\documentclass[journal=nalefd,manuscript=letter]{achemso}

\usepackage{amsmath,graphicx,lineno,siunitx,subcaption}
\usepackage[version=4]{mhchem}  

\title{Polytype-Dependent Upconversion Photoluminescence in 3R-MoS$_2$}

\author{Omri Meron}
\altaffiliation{These authors contributed equally to this work.}
\affiliation{Condensed Matter Physics Department, School of Physics and Astronomy, Tel Aviv University, Israel}
\alsoaffiliation{Center for Light-Matter Interaction, Tel Aviv University, Israel}

\author{Idan Kizel}
\email{idankizel@mail.tau.ac.il}
\altaffiliation{These authors contributed equally to this work.}
\affiliation{Condensed Matter Physics Department, School of Physics and Astronomy, Tel Aviv University, Israel}
\alsoaffiliation{Center for Light-Matter Interaction, Tel Aviv University, Israel}

\author{Dror Hershkovitz}
\affiliation{Condensed Matter Physics Department, School of Physics and Astronomy, Tel Aviv University, Israel}

\author{Youngki Yeo}
\affiliation{Condensed Matter Physics Department, School of Physics and Astronomy, Tel Aviv University, Israel}

\author{Nirmal Roy}
\affiliation{Condensed Matter Physics Department, School of Physics and Astronomy, Tel Aviv University, Israel}

\author{Wei Cao}
\affiliation{State Key Laboratory of Materials-Oriented Chemical Engineering, College of Chemical Engineering, Nanjing Tech University, China}

\author{Alon Ron}
\affiliation{Condensed Matter Physics Department, School of Physics and Astronomy, Tel Aviv University, Israel}
\alsoaffiliation{Center for Light-Matter Interaction, Tel Aviv University, Israel}

\author{Moshe Ben Shalom}
\affiliation{Condensed Matter Physics Department, School of Physics and Astronomy, Tel Aviv University, Israel}

\author{Haim Suchowski}
\affiliation{Condensed Matter Physics Department, School of Physics and Astronomy, Tel Aviv University, Israel}
\alsoaffiliation{Center for Light-Matter Interaction, Tel Aviv University, Israel}

\begin{document}

\begin{abstract}
Ferroelectric van der Waals materials offer switchable polarization states, yet optical readout of their stacking configurations remains challenging. Here, building on the resonant exciton-exciton annihilation (EEA) mechanism recently identified in 2H-phase TMDs, we report the first observation of upconversion photoluminescence (UPL) in rhombohedral MoS$_2$ and demonstrate that this many-body process is strongly polytype-dependent. Using low-temperature spectroscopy, we observe anti-Stokes emission with superlinear power dependence characteristic of EEA. Beyond serving as an accurate layer-number sensor due to discrete bandgap variations, UPL provides a sensitive probe of stacking order across thicknesses. The two neutral trilayer polytypes, which remain indistinguishable by surface potential measurements and second harmonic generation, exhibit markedly different UPL intensities. This sensitivity persists in thicker samples where multiple configurations coexist. First-principles calculations suggest that the intensity contrast originates primarily from the layer confinement of the annihilating excitons, while energy matching to the $\Gamma$ final-state manifold provides additional intensity selectivity. Power-dependent spectroscopy further disentangles two distinct annihilation channels originating from different dark exciton valleys, identified through their contrasting intensity scaling and opposite density-induced energy shifts. Crucially, the annihilation process doubles the energy separation of nearly degenerate dark excitons while converting their weak emission into bright signal, providing experimental access to valley-specific responses that are obscured in direct dark-exciton spectroscopy. Our findings demonstrate that ferroelectric configurations provide a new degree of freedom for controlling nonlinear optical processes, with implications for all-optical ferroelectric readout and electrically switchable wavelength conversion in two-dimensional materials.
\end{abstract}

\section{Keywords}
Polar van der Waals (vdW) polytypes, SlideTronics, 2D ferroelectric materials, rhombohedral molybdenum disulfide, photoluminescence, upconversion

\section{Introduction}

The coupling between ferroelectric order and optical response in two-dimensional materials opens new pathways for controlling light-matter interactions at the nanoscale. In this context, two-dimensional (2D) transition metal dichalcogenides (TMDs) are particularly promising: their reduced dimensionality and strong Coulomb interactions yield robust excitonic effects, while their structural polymorphism enables tunable electronic phases \cite{Wang2018RMP, Xiao2012PRL}. Rhombohedral (3R) polytypes of TMDs exemplify this promise, hosting interfacial ferroelectricity where interlayer sliding induces robust, switchable out-of-plane polarization, a property that underpins the emerging field of \textit{SlideTronics}\cite{vizner_stern_interfacial_2021, vizner_stern_sliding_2024}. At the same time, the strong excitonic resonances and rich manifold of bright and dark excitons in few-layer TMDs \cite{perea-causin_exciton_2022} give rise to a pronounced optical response, so that exciton recombination can be used as a sensitive optical probe of the underlying TMD properties \cite{Kizel2025ACSNano, liang_nanosecond_2025}.

Upconversion photoluminescence (UPL) in TMDs has been observed through several mechanisms, including phonon-assisted anti-Stokes emission~\cite{Jones2016NatPhys, Jadczak2019NatCommun, Dai2023LSA}, Auger-like scattering of bright excitons~\cite{Manca2017NatCommun, Paur2019NatCommun}, and plasmon-enhanced dark exciton pathways~\cite{Xu2023NatCommun}. Recently, Chen \textit{et al.} reported an efficient unique pathway: resonant exciton-exciton annihilation (EEA) of momentum-dark excitons gives rise to a pronounced UPL emission in few-layer 2H-TMDs~\cite{Chen2025NatCommun}.

However, the manifestation of such upconversion pathways in rhombohedral TMDs has not yet been studied. In these systems, distinct ferroelectric polytypes (different layer-stacking sequences with different out-of-plane polarizations) have not yet been explored in the context of UPL, leaving open the question of whether upconversion is sensitive to the underlying stacking order and thus polytype dependent. Clarifying this connection is particularly relevant because 3R-TMDs combine switchable polarization with reconfigurable domain architectures \cite{Wang2023NatMater}, offering potential for devices that integrate wavelength conversion with ferroelectric control.

Here, we report the first observation of UPL in 3R-MoS$_2$ and demonstrate strong polytype dependence of the upconversion efficiency. We focus first on rhombohedral trilayers, where three distinct polytypes coexist: ABC/CBA (flipped same polytype, carrying opposite polarizations) and centrosymmetric ABA/BAB (both with zero net polarization). While ABC and CBA are distinguishable by Kelvin probe force microscopy (KPFM), the ABA and BAB configurations remain degenerate to such measurement. By contrast, our UPL readout converts their opposite symmetry into a robust difference in upconversion efficiency, enabling reliable, large-area labeling of ABA versus BAB. We then present first-principles calculations that attribute this polytype dependence primarily to the layer confinement of the annihilating excitons, which controls the spatial overlap entering the annihilation matrix element, together with variations in the $\Gamma$-point conduction manifold governing the energy matching of the exciton-exciton annihilation process.
We further map the layer and power dependence of the upconverted emission in few-layer stacks, disentangling two distinct annihilation channels, each fed by a different dark exciton valley, through their contrasting intensity scaling and opposite density-induced shifts. Finally, we show that polytype-selective UPL persists in thicker samples where multiple polytypes coexist. This UPL-based readout complements recently established methods for mapping ferroelectric polytypes in rhombohedral TMDs \cite{Kizel2025ACSNano, liang_resolving_2025, liang_nanosecond_2025}. Together, these optical techniques provide non-invasive, large-area characterization that overcomes limitations of traditional surface potential measurements, while the coupling of UPL efficiency to switchable stacking order suggests a route toward slidetronic control of wavelength upconversion.

\section{Results and Discussion}

\subsection{Polytype-resolved upconversion in trilayer 3R-MoS$_2$}

\begin{figure}[htbp]
    \centering
    \includegraphics[width=\columnwidth]{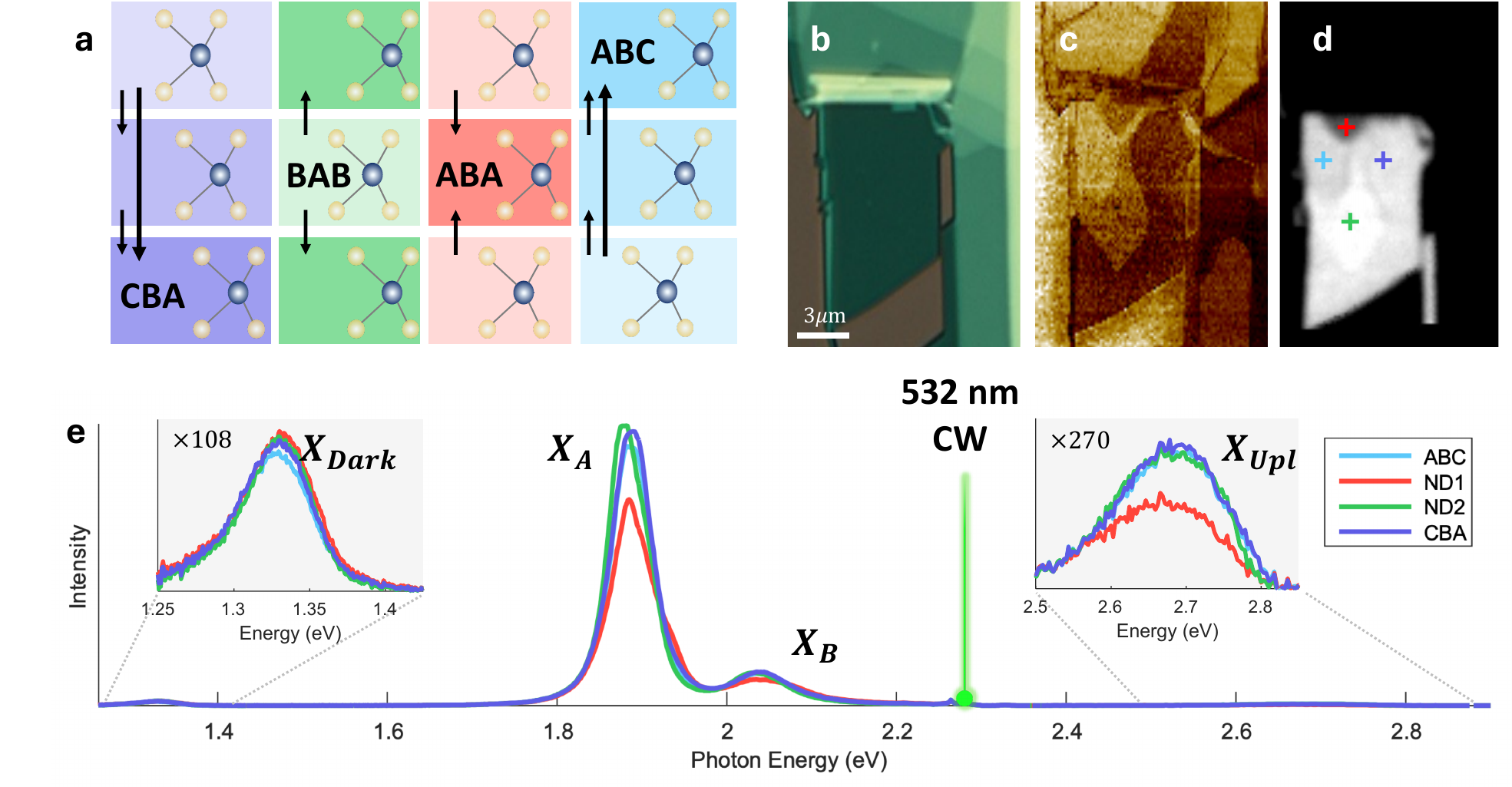}
    \caption{\textbf{Polytype-dependent UPL in trilayer 3R-MoS$_2$.}
    (\textbf{a}) Schematic of the four trilayer 3R stacking sequences (ABC, ABA, BAB, CBA), illustrating the relative layer registry and the associated out-of-plane polarization direction (arrows).
    (\textbf{b}) Optical micrograph image of the trilayer region.
    (\textbf{c}) KPFM-scan image of the same area used to delineate the domains.
    (\textbf{d}) Integrated UPL map (2.4-2.8 eV). Color markers indicating the locations assigned to each polytype (Light blue - ABC, dark blue - CBA, red - Neutral domain 1 and Green - Neutral domain 2).
    (\textbf{e}) Full PL spectra under \SI{532}{\nano\meter} excitation, comparing the four polytypes. The main panel shows the known bright excitons emission ($X_A$, $X_B$), while the insets highlight the momentum-dark exciton band ($X_{\mathrm{Dark}}$) and the upconverted emission band ($X_{\mathrm{UPL}}$).}
  \label{figure1}
\end{figure}

We begin by demonstrating polytype-dependent UPL in the simplest thickness where multiple 3R polytypes show measurable UPL. In trilayer 3R-MoS$_2$, three distinct stacking configurations are possible, ABC (and its inverted partner CBA), and two Bernal-3R variants, ABA and BAB. As sketched in Figure~\ref{figure1}a, ABC and CBA carry equal-magnitude but opposite out-of-plane polarization, whereas both ABA and BAB have zero net out-of-plane polarization.

Our trilayer device (\SI{90}{nm}-thick SiO$_2$/Si substrate) contains all three polytypes within a single contiguous flake, enabling a direct comparison under identical dielectric and excitation conditions. All UPL measurements in this work are performed at \SI{4}{K} under continuous-wave \SI{532}{nm} excitation. A conventional optical microscope image (Figure~\ref{figure1}b) allows us to determine the layer number but offers no contrast between different stacking configurations. In contrast, the Kelvin probe force microscopy (KPFM) map (Figure~\ref{figure1}c) resolves three surface-potential levels: two domains with opposite polarization contrast, ABC and CBA, and a third neutral level corresponding to the Bernal-3R family, either ABA or BAB, which cannot be distinguished.

To resolve the two Bernal-3R stackings, we perform a raster scan of the photoluminescence intensity across the same region (representative spectra in Figure ~\ref{figure1}e). The photoluminescence map effectively splits the KPFM-degenerate Bernal domains into two subdomains denoted ND1 and ND2, which exhibit markedly different upconversion efficiencies (intergration over the UPL regime in Figure~\ref{figure1}d). While the momentum-dark excitons, associated with $Q-\Gamma/K-\Gamma$ transitions, show a uniform response between ND1 and ND2 in (Figure ~\ref{figure1}e, left inset), the UPL emission is strongly suppressed in ND1 (Figure ~\ref{figure1}e, right inset). The bright A-exciton also exhibits a subtle but systematic difference between the two domains. At this stage, we use ND1 and ND2 only as experimental labels. Their assignment to ABA and BAB is developed in the following section using the layer-projected EEA model, and the A-exciton lineshapes are subsequently examined as a separate spectroscopic consistency check following the mechanism recently identified by Liang \textit{et al. ~\cite{liang_resolving_2025}}.
This discrepancy, where the upconverted signal varies despite a constant dark-exciton reservoir, strongly implies that the polytype dependence originates from the nonlinear resonant characteristics of the EEA process \cite{Chen2025NatCommun}. In the following subsection, we interpret these trends in terms of dark-exciton annihilation and the layer-resolved $\Gamma$-conduction manifold, establishing UPL as a sensitive, high-contrast probe for distinguishing these otherwise degenerate ferroelectric states.

\subsection{Band structure origin of polytype-dependent UPL}
\begin{figure*}[t]
  \centering
  \includegraphics[width=\textwidth]{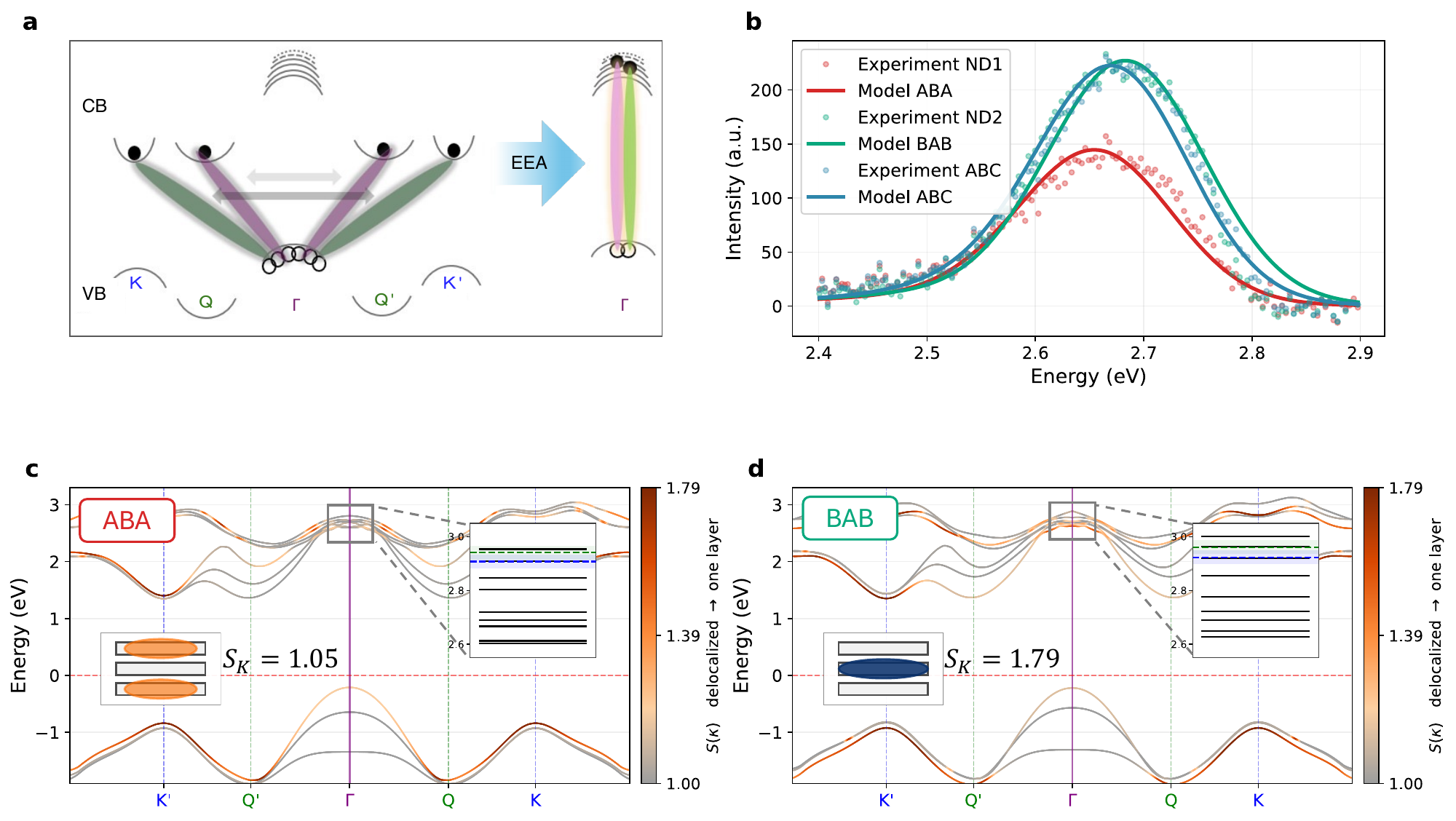}
    \caption{\textbf{Band-structure origin of polytype-dependent UPL in trilayer 3R-MoS$_2$.}
    (\textbf{a}) Schematic of the upconversion mechanism, in which two momentum-dark excitons formed between the $Q/Q'$ (and/or $K/K'$) conduction valleys and the $\Gamma$ valence band, annihilate into a bright $\Gamma-\Gamma$ exciton, enhanced by the dense cluster of conduction-band states near $\Gamma$.
    (\textbf{b})  UPL spectra of  ABC, ND1 and ND2 trilayer regions (symbols), with the fitted curves of the model (solid lines). 
    (\textbf{c},\textbf{d}) HSE band structures of trilayer ABA (c) and BAB (d) along $K-Q-\Gamma-Q'-K'$, colored according to the layer-overlap factor $S_{v} (\kappa)$. The color scale runs from $S(\kappa)=1$ for a state spread evenly over the three layers, to $S(\kappa)=1.79$ for a state confined to a single layer. The left inset in each panel shows the layer distribution of the lowest K-valley electron. The right insets shows the $\Gamma$ conduction cluster and marks the two-exciton energies $2E_{dark}^{K}$ (blue) and $2E_{dark}^{Q}$ (green), with the shaded band around the levels indicating the EEA broadening $\gamma_{EEA}$}
  \label{figure2}
\end{figure*}

The stark contrast in UPL efficiency across polytypes suggests a mechanism governed by the specific details of the rhombohedral electronic structure. While the bright A-exciton response follows the dielectric screening model proposed by Liang \textit{et al.}~\cite{liang_resolving_2025,liang_nanosecond_2025}, the UPL behavior requires a description of the many-body dynamics. Building on the resonant EEA framework recently established for 2H-TMDs~\cite{Chen2025NatCommun}, we attribute the UPL in 3R-MoS$_2$ to a process where two dark excitons with opposite momenta ($Q-\Gamma$ and $Q'-\Gamma$ or $K-\Gamma$ and $K'-\Gamma$) annihilate into a high-energy excitonic state near the $\Gamma$ point (Figure~\ref{figure2}a). Notably, in contrast to 2H-MoS$_2$ where the $Q$ valley forms the conduction band minimum for thicknesses beyond two layers, in few-layer 3R-MoS$_2$ the $Q$ and $K$ valleys are nearly degenerate. Therefore, the formation and recombination of bright excitons from the band edge $K/K'$ valleys is still quite efficient, lowering the efficiency of the EEA with respect to the 2H-MoS$_2$ case. Within this EEA framework, the upconversion efficiency depends both on the interaction matrix element and on energy and momentum matching to the available final states. Because our measurements indicate that the dark-exciton reservoir is largely insensitive to the specific rhombohedral polytype (Figure~\ref{figure1}e), the pronounced variations in UPL intensity must originate from the annihilation step.
To study this, we performed self-consistent field (SCF) calculations for the relaxed atomic models at the Heyd-Scuseria-Ernzerhof (HSE) \cite{Heyd2003JCP_HSE03, Heyd2004JCP_HSE_Solids, Heyd2004JCP_HSE_Validation, Heyd2006JCP_Erratum} screened-exchange hybrid density functional level of theory (see Methods for details). The resulting band structures layer projections (Figures~\ref{figure2}c--d) reveal the dominant stacking-sensitive contribution. The K-valley conduction electron is confined to the central layer in BAB and to the bottom layer in ABC, but is shared between the two outer layers in ABA, reducing the spatial overlap and therefore the annihilation matrix element in ABA (Figures~\ref{figure2}c--d, left insets). The band structures also reveal a "$\Gamma$ conduction cluster" consisting of a manifold of closely spaced conduction bands. Compared to BAB (and ABC), the ABA configuration exhibits an increased energy separation between this $\Gamma$ cluster and the dark-exciton annihilation energy, $2E_{\text{dark}}$ (Figures~\ref{figure2}c--d, right insets). This displacement reproduces the 25 $meV$ redshift of the UPL maximum, and provides an energy-matching contribution to the contrast. Together, the differences explain the observed UPL suppression in ABA domain despite the identical surface potential and dark-exciton population.

To test this picture quantitatively, we construct a minimal model guided by Fermi's golden rule. The transition rate is proportional to the squared matrix element times the joint density of states, evaluated at energies satisfying conservation. For EEA feeding a bright population, energy conservation requires $2E_{\mathrm{dark}} = E_{\Gamma\Gamma}^{(i)}$. Finite lifetimes and dephasing broaden this constraint: we evaluate a Lorentzian centered at zero energy mismatch, $L(E_{\Gamma\Gamma}^{(i)}-2E_{\mathrm{dark}};\,\gamma_{\mathrm{EEA}})$, which weights each channel by how well it satisfies energy matching. The subsequent radiative recombination produces a Voigt emission line, centered at the transition energy, arising from homogeneous ($\gamma_{\mathrm{hom}}$) and inhomogeneous ($\sigma_{\mathrm{inhom}}$) broadening. Retaining the layer dependence of the matrix element and writing the valley sum explicitly:
\begin{equation}
    I_{\mathrm{pred}}(E)
    =
    A
    \sum_{i}
    \sum_{v \in \{K,Q\}}
    S_{v}(\kappa)\,
    L\!\left(
    E_{\Gamma\Gamma}^{(i)}
    -
    2E_{\mathrm{dark}}^{(v)};
    \gamma_{\mathrm{EEA}}
    \right)
    V\!\left(
    E-\Delta-E_{\Gamma\Gamma}^{(i)};
    \sigma_{\mathrm{inhom}},
    \gamma_{\mathrm{hom}}
    \right)
    \label{eq:UPL_model}
\end{equation}
%
%
where $E_{\Gamma\Gamma}^{(i)}$ are DFT-computed $\Gamma\!\to\!\Gamma_i$ gaps, and $\Delta$ accounts for missing many-body corrections. The Lorentzian (centered at zero) assigns weight based on energy mismatch, while the Voigt (centered at $E_{\Gamma\Gamma}^{(i)}+\Delta$) produces the spectral line shape. 
The layer-overlap factor of the valley-$v$ exciton is $S_{v}(\kappa)\propto\sum_{\ell,\ell'}\rho_{v}(\ell)\,\rho_{v}(\ell')\,\kappa^{|\ell-\ell'|}$ (normalized to unity for a layer-delocalized exciton). Here $\rho_{v}(\ell)$ is the calculated layer density of the valley-$v$ electron, and $\kappa \in [0,1]$ is the efficiency of annihilation between excitons in adjacent layers relative to excitons in the same layer. At $\kappa = 1$ all $S_{v} = 1$ and the expression reduces to a purely energy-matched rate, while at $\kappa = 0$ it describes rigid intralayer annihilation.
A joint fit of the three trilayer spectra (ABC, and the two neutral domains) with a single shared parameter set ($\Delta = -0.48 eV,\gamma_{\mathrm{EEA}} = 0.052 eV, \sigma_{\mathrm{inhom}} = 0.062 eV, \gamma_{\mathrm{hom}} = 0.003 eV, \kappa = 0.42$) reproduces the measured spectra (Figure 2b). The band structures of the two neutral domains, with each state colored by its layer-overlap factor $S(\kappa)$ (Figure 2c,d), together with the underlying DFT layer projections (Supporting Information Figure S11), show that the K-valley conduction minimum of BAB is confined to the inner layer, $\rho_{K} = [0,1,0]$, whereas in ABA it is shared between the two outer layers, $\rho_{K} = [0.5,0,0.5]$. In the ABC (CBA) stackings (Supporting Information Figure S12) the ferroelectric potential splits the three K-valley conduction states into a layer-polarized ladder, and the lowest K electron therefore occupies a single, outer layer, $\rho_{K} = [1,0,0]$  ($\rho_{K} = [0,0,1]$) . At the fitted $\kappa$ the overlap factors at K-valley are $S_{K} = 1.05$ for ABA against $1.79$ for both BAB and ABC (CBA), while the factors at Q-valley remain close to unity ($1.02,1.14$ and $1.10$), consistent with the layer-delocalized character of the Q-valley states. Energy-matching condition alone gives $I_{ABA}/I_{BAB}  = 0.79$, while the combined model predicts $0.63$, compared with $0.68$ experimentally. We therefore identify the layer-dependent interaction overlap as the primary origin of the ABA suppression. The calculated hierarchy also provides the basis for assigning the experimental neutral domains. Because ABA is predicted to be uniquely suppressed, while BAB and ABC retain comparable higher UPL efficiencies, we assign the lower-UPL ND1 domain to ABA and the higher-UPL ND2 domain to BAB. As a separate spectroscopic consistency check, we return to the A-exciton manifolds of ND1 and ND2. The criterion established by Liang et al.~\cite{liang_resolving_2025} applies to hBN-encapsulated trilayers, where the two outer layers experience nearly equivalent dielectric environments. In our SiO$_2$/vacuum geometry, this equivalence is lifted, introducing an additional dielectric splitting between the outer-layer excitons. Combining this splitting with the layer-projected band structures predicts three underlying layer-resolved contributions for ABA and two dominant contributions for BAB. Constrained fits reproduce ND1 with the three-component ABA manifold and ND2 with the two-component BAB manifold. Because the relevant separations are smaller than the measured linewidth of approximately 55 meV, the individual components are not fully resolved in the raw spectra. The fitted spectra, component energies, and detailed analysis are presented in Supporting Information Section S4. The A-exciton lineshapes therefore provide supporting spectroscopic evidence consistent with the DFT-guided neutral domains assignment.
The A-exciton observables also constrain non-stacking origins of the contrast. Integrated over the trion region, the spectral weight of the two neutral domains is equal within 5$\%$, and the momentum-dark exciton band is indistinguishable, ruling out large doping differences~\cite{mouri_tunable_2013, mak_tightly_2012}. Since the A-exciton shifts by a few tens of $meV$ per percent strain ~\cite{Conley2013BandgapEngineering}, the 6 $meV$ difference between the composite peaks, which the layer-resolved manifold accounts for, together with the unshifted dark band and the common intrinsic linewidth of the two domains, bounds strain below the level required for the contrast.


\subsection{Layer- and power-dependent UPL in 3R-MoS$_2$}

\begin{figure}[htbp]
    \centering
    \includegraphics[width=\columnwidth]{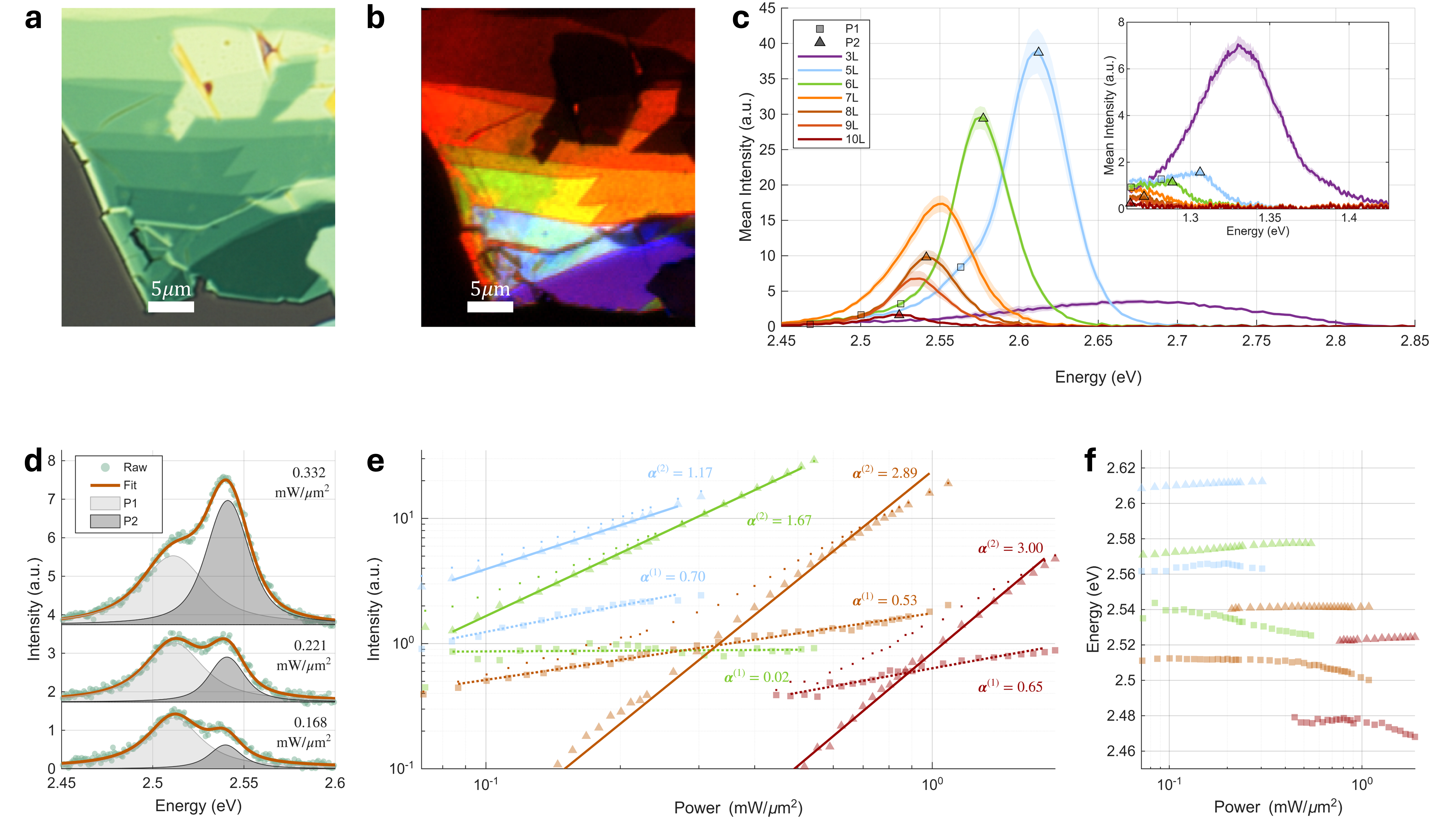} 
    \caption{\textbf{Layer- and power-dependent UPL in 3R-MoS$_2$.}
    (\textbf{a}) Optical micrograph image of the measured flake. 
    (\textbf{b}) False-color UPL-contrast image of the same area, constructed by mapping the spectrally integrated UPL intensity in three energy windows assigned to the R/G/B channels. Integration bands defined in the main text.
    (\textbf{c}) UPL spectra from $N=3$ and $N=5$--10 layers, showing a pronounced thickness dependence and two spectral components (P1, P2; peak centers marked). Inset: corresponding momentum-dark exciton spectra.
    (\textbf{d}) Example spectral fits at three excitation power densities for 6 layers (data, total fit, and two components).
    (\textbf{e}) Log--log excitation power dependence of the integrated intensity, shown separately for P1 (squares) and P2 (triangles) for each layer number, with power-law fits $I\propto P^{\alpha}$ (exponents annotated).
    (\textbf{f}) Extracted peak energies of P1 (squares) and P2 (triangles) versus excitation power.}
    \label{figure3}
\end{figure}

Having established the polytype sensitivity of UPL in trilayers, we now examine the evolution of this response across a broader range of thicknesses. We investigate a rhombohedral \ce{MoS2} flake containing regions ranging from 3 to 10 layers, as identified by white-light optical contrast (Figure~\ref{figure3}a). To visualize the spatial heterogeneity of the upconversion response, we generated a false-color ``RGB'' UPL map (Figure~\ref{figure3}b). In this map, the red, green, and blue channels correspond to spectrally integrated UPL intensities within three energy windows: R = 2.38--2.53~eV, G = 2.53--2.64~eV, and B = 2.61--2.82~eV.

The thickness-dependent UPL spectra (Figure~\ref{figure3}c) reveal a non-monotonic intensity trend. While both UPL and momentum-dark exciton peaks (inset, Figure~\ref{figure3}c) redshift with increasing layer number, consistent with the narrowing of the electronic bandgap as thickness increases~\cite{Zhao2013OriginIndirect}, the overall UPL intensity peaks at five layers before decreasing in thicker regions. This behavior can be understood through the resonance condition of the EEA process. In the rhombohedral trilayer, our DFT calculations show that the average $\Gamma$--$\Gamma$ transitions are significantly detuned from the EEA energy ($2E_{\mathrm{dark}}$). In this regime, the EEA process is less efficient, and a larger fraction of the dark-exciton reservoir decays radiatively at its own energy rather than undergoing upconversion. As the thickness increases toward five layers, the shifting band structure optimizes the energy-matching condition, maximizing UPL efficiency.

To elucidate the underlying many-body dynamics, we performed power-dependent measurements. The resulting spectra imply the UPL line shape is constructed of at least two contributions with distinguished power-law dependencies (see Figure ~\ref{figure3}d). We modeled these spectra using a decomposition consisting of a low-energy Lorentzian (light gray) and a higher-energy Fano lineshape (dark gray), both convolved with a Gaussian to account for inhomogeneous broadening (detailed fitting procedures are provided in the Supporting Information). The Fano profile of the high-energy component suggests that the emitting excitonic state interferes with an underlying continuum, likely associated with the dense $\Gamma$-conduction manifold.

The two components exhibit markedly different power-law exponents, peak energies, and density-dependent shifts (Figures 3e and 3f). These contrasts can be understood by assigning each component to a distinct dark-exciton annihilation channel. We attribute the high-energy Fano component (triangles) to annihilation of $Q$--$\Gamma$ exciton pairs and the low-energy Lorentzian component (squares) to annihilation of $K$--$\Gamma$ exciton pairs. Three observations support this assignment.
First, the low-energy component displays sub-linear power scaling while the high-energy component shows super-linear scaling that increases from $\alpha \approx 1.5$ in five-layer to $\alpha \approx 3$ in ten-layer (Figure~\ref{figure3}e). This behavior is consistent with the disparate density of states in each valley. The $K$ valleys possess only two-fold degeneracy, whereas the $Q$ valleys form a six-fold degenerate ring in momentum space and exhibit heavier effective masses due to their flatter dispersion~\cite{Jin2013PRL_DirectMeasurement, Kormanyos2015_2DMater_kp, Cheiwchanchamnangij2012PRB}. These factors suggest that the $Q$ valleys can accommodate a larger exciton population before saturating. The sub-linear scaling of the $K$--$\Gamma$ component may indicate that this reservoir approaches saturation within our measurement range, while the $Q$--$\Gamma$ channel continues to accumulate excitons and displays the super-linear dependence expected for two-body annihilation. 
The increase of the power-law exponent with thickness is consistent with the known crossover from exciton-dominated EEA in few-layer TMDs toward more bulk-like Auger recombination, for which the upconversion intensity approaches a cubic scaling~\cite{Chen2025NatCommun}.
Second, the two components exhibit opposite power-dependent energy shifts (Figure~\ref{figure3}f). The high-energy Fano component blueshifts with increasing excitation density, while the low-energy Lorentzian redshifts. This contrast reflects the different interlayer character of the two valleys. The $Q$ valleys are composed of out-of-plane orbitals with strong interlayer coupling and substantial out-of-plane dipole moments, whereas the $K$ valleys are dominated by in-plane orbitals that remain localized within individual layers~\cite{Jiang2021LSA_InterlayerExciton, Kadantsev2012SSC, Cheiwchanchamnangij2012PRB}. In rhombohedral polytypes, this $K$-valley confinement is further reinforced by symmetry~\cite{Suzuki2014NatPhys_3R}. $Q$--$\Gamma$ excitons therefore carry significant dipole moments, evidenced by the density-dependent blueshift, a hallmark of repulsive dipole--dipole interactions between excitons carrying a net out-of-plane dipole moment~\cite{Kremser2020IX, Jiang2021LSA_InterlayerExciton, federolf2025tuning, SchwandtKrause2025FerroelectricControl}. Finally, the low-energy component consistently appears $\approx$ 50~meV below the high-energy component across all measured thicknesses, in agreement with our DFT calculations (see Supporting Information). This energy separation further supports our attribution of the low-energy and high-energy components to the $K$--$\Gamma$ and $Q$--$\Gamma$ dark-exciton pathways, respectively.
This valley-resolved detection highlights a unique spectroscopic capability of UPL. While the momentum-dark exciton band appears as a single feature due to the small energy splitting of excitons from the two valleys, the annihilation process doubles the energy separation and converts the weak dark-exciton emission into a bright upconverted signal. These two factors yield well-resolved spectral components whose distinct power scaling and density-induced shifts can be tracked independently, revealing how polytype variations modulate each valley in ways that remain obscured in direct dark-exciton emission.

\subsection{Polytype-resolved upconversion in seven-layer 3R-MoS$_2$}

\begin{figure}[htbp]
    \centering
    \includegraphics[width=\columnwidth]{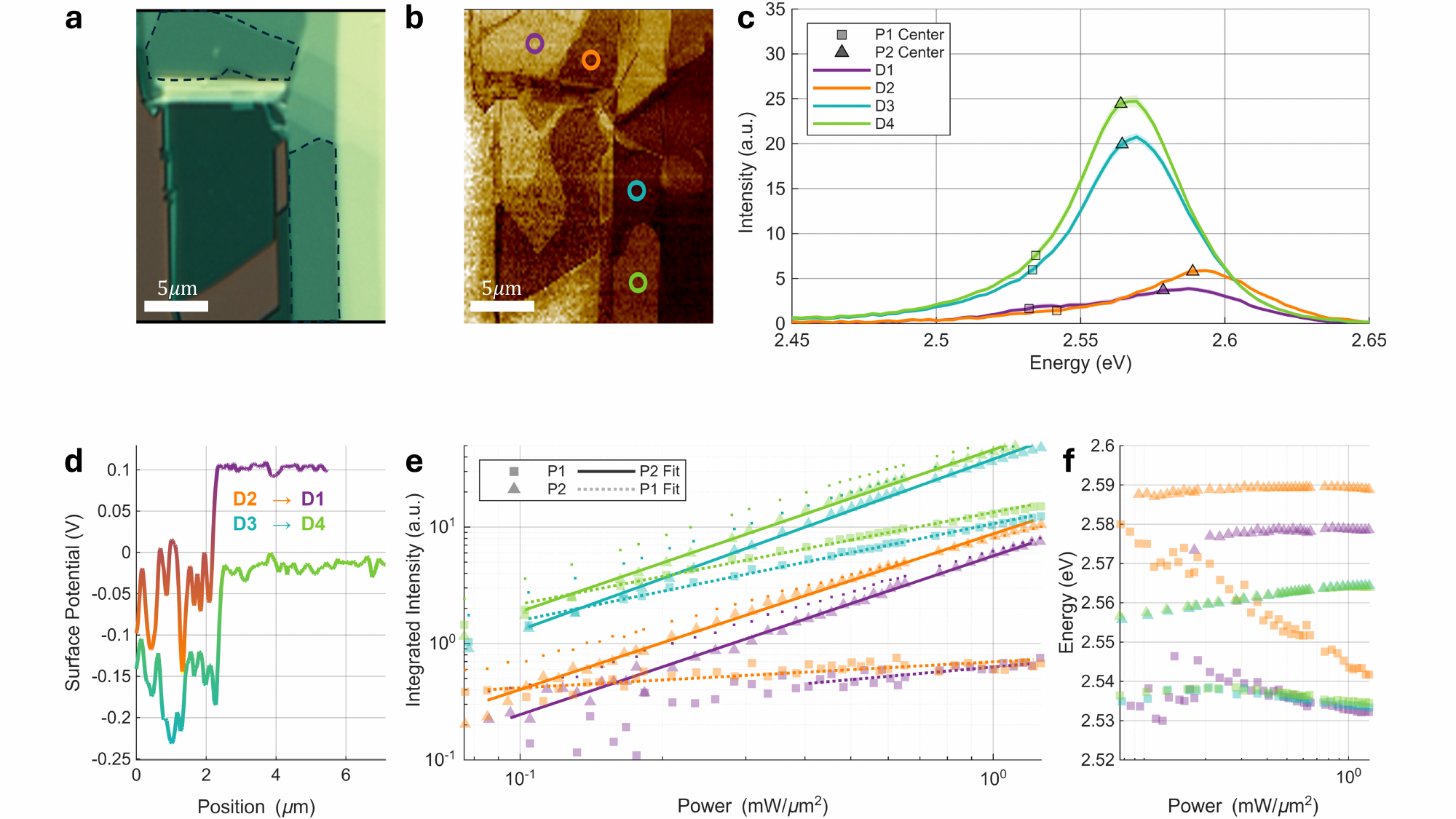} 
    \caption{\textbf{Polytype-dependent UPL in a 7L 3R-MoS$_2$ region.}
    (\textbf{a}) Optical micrograph of the measured 7L area (dashed outlines).
    (\textbf{b}) KPFM surface-potential map of the same region, with the four measurement locations (D1--D4, circles) used for the optical spectra.
    (\textbf{c}) UPL spectra acquired at D1--D4, showing strong domain-dependent intensity at fixed thickness. Symbols mark the extracted centers of the two fitted components (P1 - squares, P2 - triangles).
    (\textbf{d}) KPFM potential profiles along the straight lines connecting the circles in (\textbf{b}).
    (\textbf{e}) Log--log excitation power dependence of the integrated intensities of P1 (squares) and P2 (triangles) for each domain, with power-law fits overlaid.
    (\textbf{f}) Extracted peak energies of P1 (squares) and P2 (triangles) versus excitation power for D1--D4.}
    \label{figure4}
\end{figure}

Beyond trilayers, thicker rhombohedral stacks host a richer landscape of local stacking configurations, providing an additional handle to modulate the upconversion response within a single flake. Figure~\ref{figure4} shows a representative seven-layer region of the same rhombohedral crystal. While the optical micrograph (Figure~\ref{figure4}a) confirms a uniform thickness, the KPFM map (Figure~\ref{figure4}b) reveals multiple lateral domains separated by clear surface-potential steps (line profiles in Figure~\ref{figure4}d), consistent with distinct local polytypes at fixed layer number.

UPL spectra acquired at four marked locations (D1–D4) within this seven-layer region (Figure~\ref{figure4}c) show the same two-component line shape observed across the thickness-dependent data in Figure~\ref{figure3}, comprising a lower-energy component (P1, Lorentzian) and a higher-energy component (P2, Fano). Importantly, however, the relative amplitudes of these components vary strongly between domains under identical excitation conditions. The brightest domains (D3/D4) show substantially enhanced upconversion compared to the dimmest ones (D1/D2), demonstrating that stacking variations alone can tune the UPL efficiency even when thickness and dielectric environment are held constant.

To quantify this domain-dependent response, we analyze the power dependence of the integrated intensities of P1 and P2 using the same fitting and integration procedure applied in the multilayer data set. Both components behave similarly to the two components discussed in the previous section (Figure~\ref{figure3}), following the same two-component spectral structure and comparable power dependence across the measured range (Figure~\ref{figure4}e--f). These results show that, at fixed thickness, local stacking alone can reproduce the same spectral trends while strongly modulating the overall upconversion efficiency, making UPL a sensitive probe of polytype variations in thicker 3R-MoS$_2$.

\section{Conclusion}
We have demonstrated the first observation of upconversion photoluminescence in rhombohedral MoS$_2$, revealing strong sensitivity to stacking configuration. Distinct polytypes within the same flake exhibit markedly different UPL efficiencies, with trilayers providing a clear case where UPL discriminates between the KPFM-degenerate ABA and BAB stackings.
Our ab initio-guided analysis attributes the polytype-dependent UPL intensity primarily to the layer confinement of the annihilating excitons, which controls their spatial overlap and interaction matrix element. Stacking-dependent energy matching to the $\Gamma$ final-state manifold accounts for the observed spectral displacement and provides an additional, smaller modulation of the annihilation efficiency.
Furthermore, by combining thickness dependence with component resolved intensity power scaling and density induced energy shifts, we identify two separate EEA channels in the upconversion response and assign them to $K$--$\Gamma$ and $Q$--$\Gamma$ dark exciton reservoirs, a distinction that has not been previously established.
The annihilation process doubles the energy separation and converts weak dark-exciton emission into a bright signal, resolving contributions that otherwise overlap and providing experimental access to valley-specific responses to stacking order that cannot be obtained through direct dark-exciton spectroscopy.
The observation of analogous stacking-selective behavior in seven-layer samples, where multiple polytypes coexist, underscores the generality of this phenomenon. Beyond providing an all-optical readout for ferroelectric domains, these findings establish polytype engineering as a viable route for tailoring upconversion efficiency in 2D semiconductors, with implications for polarization-multiplexed photonics, anti-Stokes light conversion, and the integration of ferroelectric control into nanoscale optoelectronics.

\section{Methods}
\subsection{Device Fabrication}
MoS$_2$ flakes of various thicknesses were mechanically exfoliated from a bulk 3R crystal, ($"$HQ Graphene$"$), onto Gel-Pak substrates (DGL-45/17-X8). The thicknesses of 3–10L MoS$_2$ flakes were calibrated using optical intensity contrast. The MoS$_2$ flakes on Gel-Pak were transferred onto \SI{90}{nm}-thick SiO$_2$/Si substrates at \SI{100}{\celsius}.

\subsection{AFM Measurements}
Topography and side-band KPFM measurements were performed using a Park Systems NX10 AFM, employing an HQ:NSC35-Pt tip C with a resonant frequency of $\approx$ \SI{150}{kHz}. For side-band measurements, the resonant frequency of the tip was calibrated and then offset by \SI{2.5}{kHz}. The tip was excited with an AC voltage of \SI{2}{V}, and oscillation amplitudes of \SIrange{9}{15}{nm} were used for all measurements. 
The images in Fig. \ref{figure1}.c, \ref{figure4}.b were acquired using Park SmartScan software and the data analysis was performed with the Gwyddion program. The stacking order of r-MoS$_2$ domains (ABC, CBA, ABA/BAB) was determined through systematic analysis of the KPFM measurements, Following the methodology established in our previous work on polar van der Waals heterostructures \cite{vizner_stern_sliding_2024}, we identify stacking configurations based on their characteristic surface potential signatures.

\subsection{Optical characterization}
PL spectra were measured using a self-built reflection cryogenic microscope (attoDRY800 with an external objective – LUCPLFLN40X/0.6 NA). Measurements were performed with a linearly polarized \SI{532}{nm} CW laser. The PL spectra were measured using Shamrock 303i Spectrometer coupled with iDus401 CCD camera (sensor cooled via thermoelectric cooling to \SI{205}{K}). Optical images of the devices were obtained using a metallurgical microscope.

\subsection{Computational details}
To obtain the electronic structures of multilayered MoS$_2$, we performed first principles calculations based on density functional theory (DFT), by using Vienna \textit{Ab-initio} Simulation Package (VASP) \cite{Kresse1996PRB}. The structures were first relaxed, by the Perdew-Burke-Ernzerhof (PBE)\cite{Perdew1996PRL_PBE} generalized-gradient exchange-correlation density functional approximation. The interlayer interaction was realized by the Grimme-D3 dispersion correction using Becke-Johnson (BJ) damping \cite{Grimme2010JCP_DFTD3}. A plane wave energy cutoff of \SI{600}{eV} and a $k$-point mesh of 20$\times$20$\times$1 were used, with a vertical vacuum size of \SI{2}{nm} to avoid interactions between adjacent images. The simulation details were similar to the calculations reported in a former work \cite{Cao2024AdvMater_PolSaturation}, where the convergence tests were done to support the results. Following the structural relaxation, the band structure was calculated at the Heyd-Scuseria-Ernzerhof (HSE) level of theory.

\section*{Acknowledgments}
M.B.S. acknowledges funding by the European Research Council under the European Union's Horizon 2024 research and innovation program ("SlideTronics", consolidator grant agreement No. 101126257) and the Israel Science Foundation under grant No. 319/22 and 3623/21. H.S. acknowledges funding by the Israel Science Foundation (ISF) Grant No. 2312/21.

\bibliography{bibliography}

@article{Kremser2020IX,
  author  = {Kremser, Malte and Brotons-Gisbert, Mauro and Kn{\"o}rzer, Johannes and G{\"u}ckelhorn, Janine and Meyer, Moritz and Barbone, Matteo and Stier, Andreas V. and Gerardot, Brian D. and M{\"u}ller, Kai and Finley, Jonathan J.},
  title   = {Discrete interactions between a few interlayer excitons trapped at a {MoSe}$_2$--{WSe}$_2$ heterointerface},
  journal = {npj 2D Mater. Appl.},
  year    = {2020},
  volume  = {4},
  pages   = {8},
  doi     = {10.1038/s41699-020-0141-3}
}

@article{Wang2018RMP,
  author  = {Wang, Gang and Chernikov, Alexey and Glazov, Mikhail M. and Heinz, Tony F. and Marie, Xavier and Amand, Thierry and Urbaszek, Bernhard},
  title   = {Colloquium: Excitons in atomically thin transition metal dichalcogenides},
  journal = {Rev. Mod. Phys.},
  year    = {2018},
  volume  = {90},
  pages   = {021001},
  doi     = {10.1103/RevModPhys.90.021001}
}

@article{mak_tightly_2012,
	title = {Tightly bound trions in monolayer {MoS2}},
	volume = {12},
	issn = {1476-4660},
	doi = {10.1038/nmat3505},
	number = {3},
	journal = {Nature Materials},
	author = {Mak, Kin Fai and He, Keliang and Lee, Changgu and Lee, Gwan Hyoung and Hone, James and Heinz, Tony F. and Shan, Jie},
	month = dec,
	year = {2012},
	note = {Publisher: Springer Science and Business Media LLC},
	pages = {207--211},
}

@article{mouri_tunable_2013,
	title = {Tunable {Photoluminescence} of {Monolayer} {MoS2} via {Chemical} {Doping}},
	volume = {13},
	issn = {1530-6992},
	doi = {10.1021/nl403036h},
	number = {12},
	journal = {Nano Letters},
	author = {Mouri, Shinichiro and Miyauchi, Yuhei and Matsuda, Kazunari},
	month = nov,
	year = {2013},
	note = {Publisher: American Chemical Society (ACS)},
	pages = {5944--5948},
}

@article{vizner_stern_interfacial_2021,
	title = {Interfacial ferroelectricity by van der {Waals} sliding},
	volume = {372},
	issn = {1095-9203},
	doi = {10.1126/science.abe8177},
	number = {6549},
	journal = {Science},
	author = {Vizner Stern, M. and Waschitz, Y. and Cao, W. and Nevo, I. and Watanabe, K. and Taniguchi, T. and Sela, E. and Urbakh, M. and Hod, O. and Ben Shalom, M.},
	month = jun,
	year = {2021},
	note = {Publisher: American Association for the Advancement of Science (AAAS)},
	pages = {1462--1466},
}

@article{vizner_stern_sliding_2024,
	title = {Sliding van der {Waals} polytypes},
	issn = {2522-5820},
	url = {https://www.nature.com/articles/s42254-024-00781-6},
	doi = {10.1038/s42254-024-00781-6},
	language = {en},
	urldate = {2024-12-09},
	journal = {Nature Reviews Physics},
	author = {Vizner Stern, Maayan and Salleh Atri, Simon and Ben Shalom, Moshe},
	month = nov,
	year = {2024},
}

@article{perea-causin_exciton_2022,
	title = {Exciton optics, dynamics, and transport in atomically thin semiconductors},
	volume = {10},
	issn = {2166-532X},
	url = {https://pubs.aip.org/apm/article/10/10/100701/2834996/Exciton-optics-dynamics-and-transport-in},
	doi = {10.1063/5.0107665},
	language = {en},
	number = {10},
	urldate = {2025-01-02},
	journal = {APL Materials},
	author = {Perea-Causin, Raul and Erkensten, Daniel and Fitzgerald, Jamie M. and Thompson, Joshua J. P. and Rosati, Roberto and Brem, Samuel and Malic, Ermin},
	month = oct,
	year = {2022},
	pages = {100701},
}

@article{liang_resolving_2025,
	title = {Resolving polarization switching pathways of sliding ferroelectricity in trilayer {3R}-{MoS2}},
	issn = {1748-3387, 1748-3395},
	url = {https://www.nature.com/articles/s41565-025-01862-y},
	doi = {10.1038/s41565-025-01862-y},
	language = {en},
	urldate = {2025-02-05},
	journal = {Nature Nanotechnology},
	author = {Liang, Jing and Yang, Dongyang and Wu, Jingda and Xiao, Yunhuan and Watanabe, Kenji and Taniguchi, Takashi and Dadap, Jerry I. and Ye, Ziliang},
	month = feb,
	year = {2025},
}

@article{liang_nanosecond_2025,
  title       = {Nanosecond Ferroelectric Switching of Intralayer Excitons in Bilayer through Coulomb Engineering},
  author      = {Liang, Jing and Xie, Yuan and Yang, Dongyang and Guo, Shangyi and Watanabe, Kenji and Taniguchi, Takashi and Dadap, Jerry I. and Jones, David and Ye, Ziliang},
  journal     = {Physical Review X},
  year        = {2025},
  volume      = {15},
  number      = {2},
  pages       = {021081},
  month       = jun,
  doi         = {10.1103/PhysRevX.15.021081},
  note        = {Published 4 June 2025; CC BY 4.0 license},
}

@article{Heyd2003JCP_HSE03,
  author  = {Heyd, Jochen and Scuseria, Gustavo E. and Ernzerhof, Matthias},
  title   = {Hybrid functionals based on a screened Coulomb potential},
  journal = {J. Chem. Phys.},
  year    = {2003},
  volume  = {118},
  number  = {18},
  pages   = {8207--8215},
  doi     = {10.1063/1.1564060}
}

@article{Heyd2004JCP_HSE_Solids,
  author  = {Heyd, Jochen and Scuseria, Gustavo E.},
  title   = {Efficient hybrid density functional calculations in solids: {Assessment} of the {Heyd--Scuseria--Ernzerhof} screened {Coulomb} hybrid functional},
  journal = {J. Chem. Phys.},
  year    = {2004},
  volume  = {121},
  number  = {3},
  pages   = {1187--1192},
  doi     = {10.1063/1.1760074}
}

@article{Heyd2004JCP_HSE_Validation,
  author  = {Heyd, Jochen and Scuseria, Gustavo E.},
  title   = {Assessment and validation of a screened {Coulomb} hybrid density functional},
  journal = {J. Chem. Phys.},
  year    = {2004},
  volume  = {120},
  number  = {16},
  pages   = {7274--7280},
  doi     = {10.1063/1.1668634}
}

@article{Heyd2006JCP_Erratum,
  author  = {Heyd, Jochen and Scuseria, Gustavo E. and Ernzerhof, Matthias},
  title   = {Erratum: ``Hybrid functionals based on a screened {Coulomb} potential'' [{J}. {Chem}. {Phys}. 118, 8207 (2003)]},
  journal = {J. Chem. Phys.},
  year    = {2006},
  volume  = {124},
  number  = {21},
  pages   = {219906},
  doi     = {10.1063/1.2204597}
}

@article{Kresse1996PRB,
  author  = {Kresse, Georg and Furthm{\"u}ller, J{\"u}rgen},
  title   = {Efficient iterative schemes for ab initio total-energy calculations using a plane-wave basis set},
  journal = {Phys. Rev. B},
  year    = {1996},
  volume  = {54},
  number  = {16},
  pages   = {11169--11186},
  doi     = {10.1103/PhysRevB.54.11169}
}

@article{Perdew1996PRL_PBE,
  author  = {Perdew, John P. and Burke, Kieron and Ernzerhof, Matthias},
  title   = {Generalized Gradient Approximation Made Simple},
  journal = {Phys. Rev. Lett.},
  year    = {1996},
  volume  = {77},
  number  = {18},
  pages   = {3865--3868},
  doi     = {10.1103/PhysRevLett.77.3865}
}

@article{Grimme2010JCP_DFTD3,
  author  = {Grimme, Stefan and Antony, Jens and Ehrlich, Stephan and Krieg, Helge},
  title   = {A consistent and accurate ab initio parametrization of density functional dispersion correction ({DFT-D}) for the 94 elements {H--Pu}},
  journal = {J. Chem. Phys.},
  year    = {2010},
  volume  = {132},
  number  = {15},
  pages   = {154104},
  doi     = {10.1063/1.3382344}
}

@article{Cao2024AdvMater_PolSaturation,
  author  = {Cao, Wei and Deb, Swarup and Stern, Maayan Vizner and Raab, Noam and Urbakh, Michael and Hod, Oded and Kronik, Leeor and Ben Shalom, Moshe},
  title   = {Polarization Saturation in Multilayered Interfacial Ferroelectrics},
  journal = {Adv. Mater.},
  year    = {2024},
  volume  = {36},
  number  = {28},
  pages   = {e2400750},
  doi     = {10.1002/adma.202400750}
}

@article{Xiao2012PRL, author={Xiao, Di and Liu, Gui-Bin and Feng, Wanxiang and Xu, Xiaodong and Yao, Wang}, title={Coupled Spin and Valley Physics in Monolayers of {MoS}$_2$ and Other Group-{VI} Dichalcogenides}, journal={Phys. Rev. Lett.}, year={2012}, volume={108}, pages={196802}, doi={10.1103/PhysRevLett.108.196802}
}

@article{Jadczak2019NatCommun, author={Jadczak, J. and Bryja, L. and Kutrowska-Girzycka, J. and Kapuscinski, P. and Bieniek, M. and Huang, Y.-S. and Hawrylak, P.}, title={Room temperature multi-phonon upconversion photoluminescence in monolayer semiconductor {WS}$_2$}, journal={Nat. Commun.}, year={2019}, volume={10}, pages={107}, doi={10.1038/s41467-018-07994-1}
}

@article{Manca2017NatCommun, author={Manca, M. and Glazov, M. M. and Robert, C. and Cadiz, F. and Taniguchi, T. and Watanabe, K. and Courtade, E. and Amand, T. and Renucci, P. and Marie, X. and Wang, G. and Urbaszek, B.}, title={Enabling valley selective exciton scattering in monolayer {WSe}$_2$ through upconversion}, journal={Nat. Commun.}, year={2017}, volume={8}, pages={14927}, doi={10.1038/ncomms14927}
}

@article{Chen2025NatCommun,
  author  = {Chen, Yi-Hsun and Lo, Ping-Yuan and Boschen, Kyle W. and Hsu, Chih-En and Hsu, Yung-Ning and Holtzman, Luke N. and Peng, Guan-Hao and Huang, Chun-Jui and Barmak, Katayun and Holbrook, Madisen and Wang, Wei-Hua and Hone, James and Hawrylak, Pawel and Hsueh, Hung-Chung and Davis, Jeffrey A. and Cheng, Shun-Jen and Fuhrer, Michael S. and Chen, Shao-Yu},
  title   = {Efficient light upconversion via resonant exciton-exciton annihilation of dark excitons in few-layer transition metal dichalcogenides},
  journal = {Nat. Commun.},
  year    = {2025},
  volume  = {16},
  pages   = {2879},
  doi     = {10.1038/s41467-025-57991-4}
}

@article{Kizel2025ACSNano, author={Kizel, Idan and Meron, Omri and Hershkovitz, Dror and Vizner Stern, Maayan and Ron, Alon and Ben Shalom, Moshe and Suchowski, Haim}, title={Photoluminescence Detection of Polytype Polarization in r-{MoS}$_2$ Enabled by Asymmetric Dielectric Environments}, journal={ACS Nano}, year={2025}, volume={19}, number={40}, pages={35629--35637}, doi={10.1021/acsnano.5c10905}
}

@article{Wang2023NatMater, author={Wang, Chuanshou and You, Lu and Cobden, David and Wang, Junling}, title={Towards two-dimensional van der {Waals} ferroelectrics}, journal={Nat. Mater.}, year={2023}, volume={22}, pages={542--552}, doi={10.1038/s41563-022-01422-y}
}

@article{Dai2023LSA, author={Dai, Yuchen and Qi, Pengfei and Tao, Guangyi and Yao, Guangjie and Shi, Beibei and Liu, Zhixin and Liu, Zhengchang and He, Xiao and Peng, Pu and Dang, Zhibo and Zheng, Liheng and Zhang, Tianhao and Gong, Yongji and Guan, Yan and Liu, Kaihui and Fang, Zheyu}, title={Phonon-assisted upconversion in twisted two-dimensional semiconductors}, journal={Light Sci. Appl.}, year={2023}, volume={12}, pages={6}, doi={10.1038/s41377-022-01051-9}
}

@article{Xu2023NatCommun,
  author  = {Mueller, Niclas S. and Arul, Rakesh and Kang, Gyeongwon and Saunders, Ashley P. and Johnson, Amalya C. and S{\'a}nchez-Iglesias, Ana and Hu, Shu and Jakob, Lukas A. and Bar-David, Jonathan and de Nijs, Bart and Liz-Marz{\'a}n, Luis M. and Liu, Fang and Baumberg, Jeremy J.},
  title   = {Photoluminescence upconversion in monolayer {WSe}$_2$ activated by plasmonic cavities through resonant excitation of dark excitons},
  journal = {Nat. Commun.},
  year    = {2023},
  volume  = {14},
  pages   = {5707},
  doi     = {10.1038/s41467-023-41401-8}
}

@article{Paur2019NatCommun,
  author  = {Binder, J. and Howarth, J. and Withers, F. and Molas, M. R. and Taniguchi, T. and Watanabe, K. and Faugeras, C. and Wysmolek, A. and Danovich, M. and Fal'ko, V. I. and Geim, A. K. and Novoselov, K. S. and Potemski, M. and Kozikov, A.},
  title   = {Upconverted electroluminescence via {Auger} scattering of interlayer excitons in van der {Waals} heterostructures},
  journal = {Nature Communications},
  year    = {2019},
  volume  = {10},
  number  = {1},
  pages   = {2335},
  doi     = {10.1038/s41467-019-10323-9},
  url     = {https://www.nature.com/articles/s41467-019-10323-9}
}

@article{Jones2016NatPhys, author={Jones, A. M. and Yu, Hongyi and Schailbley, J. R. and Yan, Jiaqiang and Mandrus, David G. and Taniguchi, Takashi and Watanabe, Kenji and Dery, Hanan and Yao, Wang and Xu, Xiaodong}, title={Excitonic luminescence upconversion in a two-dimensional semiconductor}, journal={Nat. Phys.}, year={2016}, volume={12}, pages={323--327}, doi={10.1038/nphys3604}
}

@article{Zhao2013OriginIndirect,
  author  = {Zhao, Weijie and Ribeiro, R. M. and Toh, Minglin and Carvalho, Alexandra and Kloc, Christian and Castro Neto, A. H. and Eda, Goki},
  title   = {Origin of Indirect Optical Transitions in Few-Layer {MoS2}, {WS2}, and {WSe2}},
  journal = {Nano Lett.},
  year    = {2013},
  volume  = {13},
  number  = {11},
  pages   = {5627--5634},
  doi     = {10.1021/nl403270k}
}

@article{federolf2025tuning,
      author = {Mathias Federolf and Alexander Steinhoff and Monika Emmerling and Matthias Florian and Christian Schneider and Sven Höfling},
      title = {Tuning the non-linear interactions of hybrid interlayer excitons in bilayer MoS$_2$ via electric fields}, 
      journal = {2D Materials},
      year = {2026},
      pages = {025028},
      volume = {13},
      month = Apr,
      doi = {10.1088/2053-1583/ae566f}
}

@article{SchwandtKrause2025FerroelectricControl,
  author  = {Schwandt-Krause, Johannes and Miloudi, Mohammed El Amine and Blundo, Elena and Deb, Swarup and Heidkamp, Jan-Niklas and Watanabe, Kenji and Taniguchi, Toshio and Schwartz, Rico and Stier, A. V. and Finley, Jonathan J. and K{\"u}hn, O. and Korn, T.},
  title   = {Ferroelectric Control of Interlayer Excitons in {3R}-{MoS2}/{MoSe2} Heterostructures},
  journal = {Nano Letters},
  pages = {214–221},
  volume = {26},
  year    = {2025},
  month   = dec,
  doi     = {10.1021/acs.nanolett.5c04932},
  url     = {https://pubs.acs.org/doi/10.1021/acs.nanolett.5c04932}
}

@article{Jin2013PRL_DirectMeasurement,
  author  = {Jin, Wenjin and Yeh, Po-Chun and Zaki, Nader and Zhang, Datong and Sadowski, Jerzy T. and Al-Mahboob, Abdullah and van der Zande, Arend M. and Chenet, Daniel A. and Dadap, Jerry I. and Herman, Irving P. and Sutter, Peter and Hone, James and Osgood, Richard M.},
  title   = {Direct Measurement of the Thickness-Dependent Electronic Band Structure of {MoS$_2$} Using Angle-Resolved Photoemission Spectroscopy},
  journal = {Physical Review Letters},
  volume  = {111},
  pages   = {106801},
  year    = {2013},
  doi     = {10.1103/PhysRevLett.111.106801}
}

@article{Kormanyos2015_2DMater_kp,
  author  = {Korm\'{a}nyos, Andor and Burkard, Guido and Gmitra, Martin and Fabian, Jaroslav and Z\'{o}lyomi, Viktor and Drummond, Neil D. and Fal'ko, Vladimir},
  title   = {{$k \cdot p$} theory for two-dimensional transition metal dichalcogenide semiconductors},
  journal = {2D Materials},
  volume  = {2},
  pages   = {022001},
  year    = {2015},
  doi     = {10.1088/2053-1583/2/2/022001}
}

@article{Cheiwchanchamnangij2012PRB,
  author  = {Cheiwchanchamnangij, Tawinan and Lambrecht, Walter R. L.},
  title   = {Quasiparticle band structure calculation of monolayer, bilayer, and bulk {MoS$_2$}},
  journal = {Physical Review B},
  volume  = {85},
  pages   = {205302},
  year    = {2012},
  doi     = {10.1103/PhysRevB.85.205302}
}

@article{Jiang2021LSA_InterlayerExciton,
  author  = {Jiang, Yu and Chen, Shuo and Zheng, Weihao and Zheng, Biyuan and Pan, Anlian},
  title   = {Interlayer exciton formation, relaxation, and transport in {TMD} van der {Waals} heterostructures},
  journal = {Light: Science \& Applications},
  volume  = {10},
  pages   = {72},
  year    = {2021},
  doi     = {10.1038/s41377-021-00500-1}
}

@article{Kadantsev2012SSC,
  author  = {Kadantsev, E. S. and Hawrylak, P.},
  title   = {Electronic structure of a single {MoS$_2$} monolayer},
  journal = {Solid State Communications},
  volume  = {152},
  pages   = {909--913},
  year    = {2012},
  doi     = {10.1016/j.ssc.2012.02.005}
}

@article{Suzuki2014NatPhys_3R,
  author  = {Suzuki, Rieko and Sakano, Masahiro and Zhang, Y. J. and Akashi, Ryo and Morikawa, D. and Harasawa, A. and Yaji, K. and Kuroda, K. and Miyamoto, K. and Okuda, T. and Ishizaka, K. and Arita, R. and Iwasa, Y.},
  title   = {Valley-dependent spin polarization in bulk {MoS$_2$} with broken inversion symmetry},
  journal = {Nature Nanotechnology},
  year    = {2014},
  volume  = {9},
  number  = {8},
  pages   = {611--617},
  doi     = {10.1038/nnano.2014.148}
}

@article{Conley2013BandgapEngineering,
  author  = {Conley, Hiram J. and Wang, Bin and Ziegler, Jed I. and
             Haglund, Jr., Richard F. and Pantelides, Sokrates T. and
             Bolotin, Kirill I.},
  title   = {Bandgap Engineering of Strained Monolayer and Bilayer {MoS$_2$}},
  journal = {Nano Letters},
  year    = {2013},
  volume  = {13},
  number  = {8},
  pages   = {3626--3630},
  doi     = {10.1021/nl4014748},
  url     = {https://doi.org/10.1021/nl4014748}
}

\end{document}